\documentclass[11pt]{article}
\usepackage{epstopdf}                                                             
\usepackage[a4paper,hmarginratio=1:1,vmarginratio=2:3,totalwidth=15.4cm,totalheight=22.6cm]{geometry}
\usepackage{bm,epstopdf,epsfig,amsmath,amssymb,amsfonts,colordvi,latexsym,wrapfig,comment,cancel,verbatim,slashed}
\usepackage{epsfig,bbm}
\usepackage{graphicx,graphics}
\usepackage[font=md,captionskip=8pt]{subfig}
\usepackage[usenames,dvipsnames]{color}
\usepackage[noadjust]{cite}
\usepackage{xcolor} 
\usepackage[utf8]{inputenc}
\usepackage{setspace}

\DeclareMathOperator{\Arcsinh}{arcsinh}
\usepackage[utf8]{inputenc}
\newcommand{\q}{\alpha}

\newcommand{\sg}{\sqrt{g}}    

\newcommand{\w}{\omega}

\allowdisplaybreaks
\usepackage{amsmath}

\newcommand{\cA}{{\cal A}}

\newcommand{\cE}{{\cal E}}
\newcommand{\cF}{{\cal F}}

\newcommand{\cK}{{\cal K}}
\newcommand{\cL}{{\cal L}}

\newcommand{\cO}{{\cal O}}

\newcommand{\cV}{{\mathcal V}}

\newcommand{\ra}{\rightarrow}

\newcommand{\be}{\begin{equation}}
\newcommand{\ee}{\end{equation}}
\newcommand{\bea}{\begin{eqnarray}}
\newcommand{\eea}{\end{eqnarray}}
\newcommand{\Ra}{\Rightarrow}

\newcommand{\baa}{\begin{array}}
\newcommand{\eaa}{\end{array}}

 \long\def\symbolfootnote[#1]#2{\begingroup
\def\thefootnote{\fnsymbol{footnote}}\footnote[#1]{#2}\endgroup}

\usepackage{ulem}

\begin{document} 
\begin{flushright}
\end{flushright}
\bigskip\medskip
\thispagestyle{empty}
\vspace{1cm}

\begin{center}
\vspace{1.cm}
{\Large \bf   Standard Model in conformal geometry:
  \bigskip

  local vs gauged scale invariance} 

 \vspace{1cm}
 
 {\bf D. M. Ghilencea}$^{\,\,a}$
 \symbolfootnote[1]{E-mail: Dumitru.Ghilencea@cern.ch}
 and
 {\bf C. T. Hill}$^{\,\,b}$
 \symbolfootnote[2]{E-mail: Hill@fnal.gov}
 
\bigskip

$^a$ {\small Department of Theoretical Physics, National Institute of Physics
 \smallskip 

 and  Nuclear Engineering (IFIN), Bucharest, 077125 Romania}

\bigskip
$^b$ {\small Fermi National Accelerator Laboratory\smallskip

 P.O. Box 500, Batavia, Illinois 60510, USA}

\end{center}

\medskip

\begin{abstract}\noindent
  We discuss comparatively local versus gauged Weyl symmetry beyond Standard Model (SM)
  and  Einstein gravity and their geometric interpretation. The SM and Einstein gravity admit
  a natural embedding in Weyl {\it integrable}  geometry
  which is a special limit of  Weyl conformal (non-metric) geometry.
  The theory has a  {\it local} Weyl scale symmetry but no associated gauge boson.
  Unlike previous models with such symmetry,
  this embedding is truly minimal i.e. with  no  additional fields beyond  SM
  and underlying geometry.  This theory is compared  to a similar minimal embedding of SM
  and Einstein gravity in Weyl conformal  geometry (SMW) which has a
  full {\it gauged} scale invariance, with  an associated Weyl gauge boson.
  At large field values, both  theories give realistic, Starobinsky-Higgs like inflation.
  The broken phase of the current model  is the  decoupling limit of  the massive Weyl
  gauge boson of the broken phase of SMW,  while   the   local scale
  symmetry of the current model is part  of the larger gauged scale symmetry of SMW.
  Hence, the current theory has a gauge embedding in  SMW.
  Unlike in the SMW, we note that in models with local scale symmetry
  the associated current is trivial, which is a concern for the  physical meaning 
  of this symmetry.   Therefore, the SMW is a more fundamental UV completion  of SM
  in a full {\it gauge theory} of scale invariance that generates
  Einstein gravity in the (spontaneously) broken phase, as an effective theory.
\end{abstract}

\newpage

\section{Introduction}

The Standard  Model (SM) with the Higgs mass parameter set to zero has a classical global
scale symmetry. This could indicate that  this symmetry is more fundamental and may
play a role in physics beyond SM \cite{Bardeen}, including gravity.
When this  symmetry is implemented in physics beyond the SM \`a la Brans-Dicke \cite{BD},
(e.g. \cite{Ga,Fe1,MS1,MS2}) it  can be  broken spontaneously and there is a
Nambu-Goldstone boson (dilaton $\phi$) which 
generates the Planck scale $M_p\propto\langle\phi^2\rangle$. The  constant vev
$\langle\phi\rangle$ can be the result of a cosmological evolution in a Friedmann-Robertson-Walker
universe \cite{Fe2}. Associated to this symmetry there is a conserved global Weyl current
$K_\mu\!\propto\! \partial_\mu \phi^2$ \cite{Ga,Fe1}. Since the dilaton  decouples there are
no fifth force constraints \cite{Fe3}.
While global symmetries are usually  broken 
by gravitational effects (black holes) \cite{Kallosh,Banks,Witten},
this might not  apply to global scale symmetry which is not compact.

Alternatively, one could  consider SM with a {\it local} (Weyl) scale symmetry or
conformal symmetry \cite{monin,luty}.
Interesting theories of the  SM  endowed  with this symmetry,
embedded in a Riemannian geometry, are found in \cite{Turok,tHooft}.
The action is linear-only in the scalar curvature and  Einstein gravity
is recovered in the broken phase. To implement this symmetry and
to  generate spontaneously the Planck scale and the Einstein term,
the addition of a new field (dilaton) to the  SM spectrum and Einstein gravity
is necessary.
But unlike in the global case, the associated current to this symmetry
is trivial \cite{J1,J2}, raising concerns  about its  physical meaning
or about its geometric interpretation \cite{Ohanian} demanded for
a theory including gravity.
Further, the new field that generates  the Planck scale is usually a ghost
(has a negative kinetic term)  which some may regard as a  concern; however,
this field is not physical since it decouples  in the (physical) Einstein frame.
Questions remain however: can we do better e.g. can we avoid adding ``by hand'' the otherwise
necessary scalar field (ghost/dilaton)? is it possible to avoid  the trivial current? 
can we implement this symmetry in SM embedded in more general  gravity theories
quadratic in curvature (rather than linear)? 

These questions are  elegantly  answered by applying the successful
gauge invariance principle to gravity, just like in the SM. One should
thus consider the full, {\it gauged} Weyl scale symmetry \cite{Weyl1,Weyl2,Weyl3} in
the physics beyond the SM and Einstein gravity. 
This means that the above local Weyl symmetry is ``completed''  by the presence of an associated
Weyl gauge boson $\omega_\mu$. We thus  distinguish here between 
local versus  gauged Weyl symmetry, as dictated by the absence or presence of
$\omega_\mu$, respectively. In fact in the former case $\w_\mu$ is simply a ``pure gauge'' field.
When we consider theories with a gauged Weyl symmetry, it can be shown that there 
exists a non-trivial conserved current $J_\mu$ associated to this symmetry
\cite{SMW,Ghilen1,Palatini}. As expected, $J_\mu$ is now just a Weyl-covariant version of
the aforementioned current $K_\mu$ of global scale symmetry,
hence $J_\mu \propto(\partial_\mu -\alpha\,\omega_\mu )\phi^2$, where $\alpha$ is
the Weyl gauge coupling.

Early  models with gauged Weyl  symmetry were  linear-only in scalar curvature
($\tilde R$) \cite{Dirac,Smolin,Kugo,Cheng,Fulton,
  Nishino,Tann,Ohanian,Moffat1,Moffat2,Guendelman,ghilen,pp1,pp2,pp3,pp4,pp5,Aluri,Huang}
hence they still  needed a scalar field be added ``by hand'' to implement this symmetry and
 generate  Einstein action and 
Planck scale $M_p\sim\langle\phi\rangle$  from a term $\phi^2 \tilde R$.
Notably, in \cite{Smolin} the Weyl gauge field was shown to
become massive after ``eating'' the dilaton $\phi$ and decouple.

General actions with gauged Weyl symmetry that were {\it quadratic}  in the  curvature
were studied in  \cite{Ghilen1,Palatini},  with an action as given by the original
gravity theory of Weyl \cite{Weyl1,Weyl2,Weyl3} where this symmetry was first introduced. 
The SM with this gauged Weyl symmetry  was also studied in this framework 
in \cite{SMW} and hereafter this is called SMW. The results of  \cite{Ghilen1,Palatini,SMW}
show that, even in the absence of matter (SM, etc), this symmetry is spontaneously
broken in a Stueckelberg mechanism  in which the Weyl gauge boson $\w_\mu$
``eats'' the would-be Goldstone (dilaton $\phi$) and becomes massive.
Hence there is no dilaton (ghost) in the spectrum, as expected in a spontaneously
broken gauge theory. With $\w_\mu$  now massive, it decouples and Einstein gravity is
obtained as a ``low energy'' effective theory and broken phase of a quadratic gravity
with gauged Weyl symmetry. 
This is important because we have: a)~an embedding of Einstein gravity into a
(quantum) gauge theory of dilatations (which  is anomaly-free \cite{WG});
b) a  general quadratic gravity action and c)  the dilaton (Stueckelberg field)
is not added ``by hand'' to implement this symmetry (as  in theories linear in $R$),
but is  part of the underlying geometry (see later).
For more details on this and SMW see \cite{SMW,Ghilen1,Palatini,WG}.

There remains unexplored  the special case of local (rather than gauged) Weyl symmetry
for the SM  embedded in quadratic gravity actions  with this symmetry.  Given the
large interest in this symmetry 
e.g. \cite{Turok,tHooft,wig1,wig3,wig4,wig5,wig6} (possibly because its underlying
geometry is metric, unlike for the  SMW), here we  study this special  case.
We are  interested in  its {\it exact} relation to  the  SM embedded
in quadratic gravity with  gauged Weyl symmetry (SMW) \cite{SMW}.
The local Weyl symmetry of the current model
is  part of the larger gauged Weyl symmetry of SMW, while its broken phase
will be  shown to be the  decoupling limit of the massive $\w_\mu$ of the broken
phase of SMW\footnote{Correspondingly, the metric/integrable Weyl geometry underlying 
  the local Weyl symmetry model is a special limit of the non-metric
  Weyl  geometry underlying the  SMW \cite{non-metricity}. In the former, $\w_\mu$
  is ``pure gauge''.}.
Hence the models are closely related, as we detail. Both models have a similar scalar potential and share
realistic inflation predictions \cite{WI3,WI1,WI2}: their tensor-to-scalar ratio $r$ is
bounded from above by that of the  Starobinsky model.

Both models are  naturally formulated in
Weyl conformal  geometry \cite{Weyl1,Weyl2,Weyl3}  because this geometry
has  Weyl symmetry built in, i.e.  the  connection  has this symmetry.
No  knowledge of Weyl geometry is required here -  we  keep its use to a  minimum and  emphasize
the more familiar gauge theory perspective for this symmetry.
The corresponding dilaton/would-be-Goldstone field  that generates all the scales
(Planck, cosmological constant, $m_\w$) of both models  is  part of the Weyl-invariant quadratic
term $\sqrt{g}\,\tilde R^2$ in the action -- hence it  has a geometric origin and so do all mass scales
it generates  \cite{non-metricity}. With no new fields added ``by hand'', our approach 
to endow the  SM with  local or gauged Weyl symmetry by embedding  it in conformal
geometry is natural and truly minimal (which is not the case of models built in (pseudo)Riemannian geometry).

The ultimate question for the  SM in quadratic gravity
with local Weyl symmetry is  whether it can be a UV  complete theory.
We show that the current associated to this symmetry is vanishing, similar
to previous models with this symmetry in Riemannian geometry (that were linear-only
in $R$). This result   justifies in our opinion the need for a full,  gauged Weyl
symmetry as in the  SMW where the associated current is non-trivial and conserved.
Our comparative study concludes  that the SMW  is a more fundamental
theory, possibly renormalizable\footnote{
  Simple  power-counting and symmetry arguments support this view,
  see \cite{SMW} (section 3), \cite{non-metricity} (section~4.6)},
that acts  as a UV completion of the SM with local scale symmetry,
giving a full gauge theory of scale invariance of  both Einstein gravity and SM.

The plan of the paper is as follows: in section~\ref{s2} we review
how local Weyl symmetry is implemented in Weyl geometry and
how Einstein gravity emerges, without new scalar/matter fields added ``by hand'' to this end.
In Section~\ref{SM}  we consider  SM with local Weyl
symmetry versus  SM with gauged Weyl symmetry  - in the former, $\omega_\mu$ is simply a 
``pure gauge'' field (non-dynamical). We compare the results of the two models, their similarities
and differences in the action and predictions for inflation. Our  conclusions are followed by an
Appendix with details of the Lagrangians of the models, equations of motions and other information.

\section{Local vs gauged  Weyl symmetry  \&  Einstein gravity}\label{s2}

In this section we briefly review comparatively the local and gauged Weyl symmetry,
their geometric  interpretation  in Weyl geometry and how Einstein gravity
is recovered in the broken phase from quadratic gravity with such symmetry,
in the absence of matter.

\subsection{Weyl conformal geometry and its gravity}

Let us first define the symmetry.
A {\it local} Weyl symmetry is the invariance of an action
under transformation $\Sigma(x)$ of (\ref{WG}) below together with (\ref{WG2}) if 
scalars $\phi$ and fermions $\psi$ are also
present\footnote{Our conventions: metric $(+,-,-,-)$, $g\!=\!\vert \det g_{\mu\nu}\vert$,
  $R^\lambda_{\mu\nu\sigma}
\! =\partial_\nu \Gamma_{\mu\sigma}^\lambda-\partial_\sigma \Gamma_{\mu\nu}^\lambda+...$,
$R_{\mu\nu}\!=\! R^\lambda_{\mu\lambda\nu}$,
$R\!=\!g^{\mu\nu} R_{\mu\nu}$.}
\bea\label{WG}
&&\hat g_{\mu\nu}=\Sigma^q \,g_{\mu\nu}, \quad
\sqrt{\hat g}=\Sigma^{2\,q}\sqrt{g},\quad
\hat e^a_\mu=\Sigma^{q/2}\, e^a_\mu,\quad
\\[5pt]
&&\,\,\,\,\hat\phi=\Sigma^{-q/2}\,\phi,\quad
\hat\psi=\Sigma^{-3 q/4} \psi.\quad
\label{WG2}
\eea

\medskip\noindent
The {\it gauged} Weyl symmetry (or Weyl gauge symmetry)
is defined as invariance of the action under (\ref{WG}), (\ref{WG2})
and (\ref{WGg}) below; eq.(\ref{WGg})  is the transformation  of an associated Weyl gauge field $\omega_\mu$
naturally expected from gauge invariance principle, if scale symmetry is indeed gauged:
\bea\label{WGg}
\hat\w_\mu=\w_\mu-\frac{1}{\alpha}\,\partial_\mu\ln\Sigma.\quad
\eea

Above we denoted by $q$ the Weyl charge of the metric and $\Sigma=\Sigma(x)>0$ is a
positive definite function, $\alpha$ is the Weyl gauge coupling and $e_\mu^a$ is the tetrad.
Since this is a scale symmetry, not an internal symmetry, there is no $i$ factor in
these transformations.  In this work we set $q\!=\!1$ and the general case is restored by simply rescaling
$\alpha\ra \alpha q$ in the results. So  local Weyl symmetry
 implicitly assumes that  $\omega_\mu=0$ or that it  is a ``pure gauge'' field.

With this notation, by Weyl geometry\footnote{For a very brief guide into Weyl conformal
  geometry  see Appendix A in \cite{SMW}.} we mean a geometry  invariant
under the above transformations  i.e. it is defined by  classes of equivalence
of the metric  and Weyl field, $(g_{\mu\nu}, \w_\mu)$,  related by (\ref{WG}), (\ref{WGg}). 
The definition of Weyl geometry is completed by eq.(\ref{wge}) below
for the Weyl connection $\tilde \Gamma$, which states that this geometry is non-metric
($\tilde \nabla_\mu g_{\alpha\beta}\!\not=\!0$):
\medskip
\bea\label{wge}
\tilde\nabla_\mu g_{\alpha\beta}=-\alpha\,q\, \w_\mu \,g_{\alpha\beta},\quad\textrm{where}\quad
\tilde\nabla_\mu g_{\alpha\beta}\equiv \partial_\mu g_{\alpha\beta}
-\tilde\Gamma_{\alpha\mu}^\rho g_{\rho\beta}
-\tilde\Gamma_{\beta\mu}^\rho g_{\rho\alpha}.
\eea

\medskip\noindent
The action of $\tilde\nabla$ may be re-written in a ``metric geometry''
format by a substitution\footnote{Re-writing eq.(\ref{wge}) as in (\ref{m}) means the theory
  becomes metric with respect to a new differential operator (see \cite{Condeescu} for a discussion).
  As explained
  in Appendix~A, while in Riemannian geometry the length of a vector is constant
  under parallel transport, eq.(\ref{wge}) of Weyl geometry means that in general only  the relative length
 (ratio)  of arbitrary two vectors (of same Weyl charge) remains invariant.
 This is consistent with the argument that physics should be independent
 of the units of length. Actually,  the norm of the vector is itself invariant
 provided that its tangent-space counterpart has vanishing Weyl charge (is invariant), see Appendix~A.
  Finally, if the Weyl gauge boson is ``pure gauge'' the length of any vector remains constant
under parallel transport  (the length curvature tensor then vanishes and the geometry is called integrable).
 }
\medskip
 \bea\label{m}
\tilde\nabla^\prime _\mu g_{\alpha\beta}=0,
\qquad \tilde\nabla^\prime\equiv \tilde\nabla\Big\vert_{\partial_\mu\ra\partial_\mu +
  \alpha\, q\, \w_\mu}.
\eea
which means that $\partial_\mu$ acting on the metric $g_{\alpha\beta}$
is replaced by its Weyl covariant derivative, as expected from  a gauge invariance perspective.
Then  $\tilde\Gamma$ can be found from the Levi-Civita connection ($\Gamma$) by the
same substitution or by direct calculation from (\ref{m}), giving (with $q=1$)
\bea\label{tGammap}
\tilde \Gamma_{\mu\nu}^\lambda=\Gamma_{\mu\nu}^\lambda\Big\vert_{\partial_\mu\ra\partial_\mu +\alpha\, \w_\mu}=
\Gamma_{\mu\nu}^\lambda+\frac{\alpha}{2}\, \,\Big[\delta_\mu^\lambda\,\, \w_\nu +\delta_\nu^\lambda\,\, \w_\mu
- g_{\mu\nu} \,\w^\lambda\Big],
\eea
where
\bea
\Gamma_{\mu\nu}^\lambda=(1/2)
\,g^{\lambda\alpha}(\partial_\mu g_{\alpha\nu} +\partial_\nu g_{\alpha\mu}-
\partial_\alpha g_{\mu\nu}).
\eea

\medskip\noindent
So the Weyl connection $\tilde\Gamma$ is completely determined by the metric and $\w_\mu$.
Unlike $\Gamma$, $\tilde \Gamma$ is gauge invariant i.e. is invariant
under (\ref{WG}), (\ref{WGg}): the transformation
of the metric (in $\Gamma$) is compensated by that of $\w_\mu$, leaving $\tilde \Gamma$ invariant.
Thus,  an action 
invariant under (\ref{WG}), (\ref{WG2}), (\ref{WGg}), also has its
underlying geometry (i.e. connection $\tilde\Gamma$) invariant.
This is very important for the consistency of the action, because the underlying geometry ``is''
physics\footnote{i.e. $\omega_\mu$ which defines the connection
  is physical, being dynamical, see later.}
so one cannot ``separate''  it (e.g. $\omega_\mu$) from the action itself and so
it should  have the same symmetry, too. Weyl geometry enables this  by construction\footnote{
  This is unlike in theories with local scale symmetry in Riemannian geometry
where $\Gamma$ is not invariant.} and this  has implications for the scalar curvature.
With (\ref{tGammap}) one computes the scalar curvature $\tilde R$ of Weyl geometry by usual
formula in terms of the connection and finds
\bea\label{tildeR}
\tilde R=R-3\,\q\,\nabla_\mu\w^\mu-\frac32 \q^2\,\w_\mu \,\w^\mu,
\\
\tilde C_{\mu\nu\rho\sigma}^2=C_{\mu\nu\rho\sigma}^2+\frac32 \,\alpha^2 \,F_{\mu\nu}^2\
\eea
Here the  rhs  is in a Riemannian notation, so
$\nabla_\mu\w^\lambda=\partial_\mu \w^\lambda+\Gamma^\lambda_{\mu\rho}\,\w^\rho$.
Here
$\tilde R=R(\tilde \Gamma,g)$  is the scalar curvature of Weyl geometry,
$\tilde R=g^{\mu\nu} \tilde R_{\mu\nu}(\tilde\Gamma)$, 
defined by $\tilde \Gamma$ while $R$ is the scalar curvature of Riemannian case.
Also $\tilde C_{\mu\nu\rho\sigma}$ ($C_{\mu\nu\rho\sigma}$)
is the Weyl tensor of Weyl (Riemannian) geometry, respectively.
What is important here is that   $\tilde R$ transforms covariantly under  (\ref{WG}), (\ref{WGg}),
$\hat{\tilde R}=\tilde R/\Sigma$,  unlike in the
Riemannian case. This  is relevant  for constructing individual Lagrangian terms invariant
under (\ref{WG}), (\ref{WG2}), (\ref{WGg})
using general covariance and gauge principles.

With the above introduction, we can write the  most general gravity action
invariant under (\ref{WG}), (\ref{WGg}) and defined by  Weyl geometry \cite{Weyl1,Weyl2,Weyl3} 
\medskip
\bea\label{L1}
\cL_1=\sqrt{g}\,\,\Big\{\frac{1}{4!\,\xi^2} \tilde R^2-\frac{1}{\eta^2}\, \tilde C_{\mu\nu\rho\sigma}^2
-\frac{1}{4}\,F_{\mu\nu}^2\Big\},
\qquad 0<\xi,\,\eta <1.
\eea

\medskip\noindent
which is Weyl's original action.
Here $F_{\mu\nu}$ is the field strength of the Weyl gauge field. Since
$\tilde \Gamma_{\mu\nu}^\lambda=\tilde \Gamma_{\nu\mu}^\lambda$, then
$F_{\mu\nu}=
\tilde\nabla_\mu \w_\nu-\tilde\nabla_\nu \w_\mu
=\partial_\mu\w_\nu-\partial_\nu \w_\mu$, similar to $F_{\mu\nu}$ in  flat space-time.

Even in the absence of matter (as above), $\cL_1$  is a realistic gauge theory of dilatations,
since it  has spontaneous breaking (Stueckelberg mechanism) in which $\omega_\mu$ becomes massive,
so it decouples and  Einstein gravity is obtained in the broken phase  as shown in
\cite{Ghilen1}. For the analysis of  $\cL_1$  see \cite{Ghilen1,SMW}; for convenience
we present its equations of motion in  Appendix~\ref{BB}. For the study of the  $C_{\mu\nu\rho\sigma}^2$
term which is largely spectator in our analysis see \cite{Mannheim,Kaku}.

\subsection{Weyl integrable geometry and Einstein gravity}

Here we are mostly concerned  with the  special case of local scale symmetry,
that corresponds to the case when $\w_\mu$ vanishes or is ``pure gauge''. In such case, obviously
\bea\label{FF}
F_{\mu\nu}=0.
\eea
In fact this is a condition for the underlying geometry (since $\w_\mu$ is geometric in origin)
and actually defines a special limit of Weyl geometry,  called {\it integrable} geometry.
What  eq.(\ref{FF})  means is  that the length curvature tensor (represented by
$F_{\mu\nu}$) of Weyl geometry
vanishes. Physically this means there is no kinetic term for $\w_\mu$.
An example of such  a theory  is conformal gravity \cite{Kaku} (where
a kinetic term  for gauged dilatations is not present).
The vanishing of $F$  implies  (assuming no topological restrictions,
simply connected smooth manifold),  that
$\w_\mu\propto \partial_\mu$(scalar field), so $\omega_\mu$ is ``pure gauge''.
For a  suitable $\Sigma$ one can then obtain  $\hat\omega_\mu=0$
which means $\tilde\Gamma=\Gamma$,   $\tilde\nabla_\mu g_{\alpha\beta}=0$
and the geometry is metric. Weyl integrable geometry is thus associated
to local Weyl symmetry while the gauged Weyl symmetry is associated to
general Weyl geometry; non-metricity discussed earlier is  necessary to have a true gauge
theory (i.e. $\omega_\mu$ dynamical).

With (\ref{FF}), the Lagrangian $\cL_1$ becomes
\smallskip
\bea\label{L1int}
\cL_1=\sqrt{g}\,\,\Big\{\frac{1}{4!\,\xi^2} \tilde R^2-\frac{1}{\eta^2}\, C_{\mu\nu\rho\sigma}^2
\Big\},
\qquad 0<\xi,\eta <1.
\eea

\medskip\noindent
Since according to (\ref{tildeR}), $\tilde R^2$ contains Riemannian
$R^2$, $\cL_1$ is  a  higher derivative theory that propagates a
spin-zero mode (from  $R^2$), in addition to the graviton.
It is easy to ``unfold'' this higher derivative theory into a second order one
and extract this spin-zero mode
from $\tilde R^2$. To this end, replace
$\tilde R^2\ra - 2 \phi^2 \tilde R-\phi^4$ in $\cL_1$, where $\phi$ is a scalar field \cite{Ghilen1}
\medskip
\bea\label{oo}
\cL_1&=&\sqrt{g}\,\Big\{\frac{1}{4!\,\xi^2}\,\Big[- 2 \phi^2 \tilde R-\phi^4\Big]
-\frac{1}{\eta^2}\, C_{\mu\nu\rho\sigma}^2\Big\}.
\\
&=&\sqrt{g}\,\Big\{\frac{1}{4!\,\xi^2}\,\Big[- 2 \phi^2 \,\Big( R-3\q \,\nabla_\mu\w^\mu -\frac32\,
\q^2 \w_\mu \w^\mu\Big)-\phi^4\Big]
-\frac{1}{\eta^2}\, C_{\mu\nu\rho\sigma}^2\Big\}.\label{ooo}
\eea

\medskip\noindent
where we used (\ref{tildeR}). The equation of motion of $\phi$ has
solution $\phi^2\!=-\tilde R$ which replaced in $\cL_1$ recovers  (\ref{L1})
which is thus classically equivalent to (\ref{oo}). The equation of motion of $\w_\mu$ is
\bea\label{omu}
\w_\mu=\frac{1}{\q}\, \partial_\mu\ln \phi^2.
\eea

\medskip\noindent
Using (\ref{omu}) back in $\cL_1$
\medskip
\bea\label{RMW}
\cL_1=\sqrt{g}\,
\Big\{-
\frac{1}{2\,\xi^2}\,
\Big[\,\frac16 \,\phi^2\,R + g^{\mu\nu} \partial_\mu \phi\,\partial_\nu \phi \Big]-
\frac{1}{4!\,\xi^2}\,\phi^4-\frac{1}{\eta^2}\,C_{\mu\nu\rho\sigma}^2\Big\}.
\eea

\medskip\noindent
This is the gravity action  in the absence of matter which has local Weyl invariance left.
  Note from (\ref{RMW}) that the field  $\phi$ is dynamical.
  Eq.(\ref{tildeR}) then gives
\medskip
\bea
\tilde R=R- 6 \nabla_\mu\nabla^\mu\,\ln\phi- 6 (\partial_\mu\ln\phi) (\partial^\mu\ln\phi).
\eea

\medskip\noindent
which relates the Weyl scalar curvature ($\tilde R$) to its Riemannian counterpart ($R$). 
Note that in $\cL_1$ the field  $\phi$   was {\it not} added by hand
but  is actually part of  the underlying  geometry \cite{non-metricity};
$\phi$ comes from the spin zero mode propagated by the $\tilde R^2$ term in the original action
(which contains $R^2$) and is  part of Weyl scalar curvature.
This situation is improved compared to  models with local scale symmetry which are
linear-only in the curvature, built in Weyl, Riemannian or some other geometry: in these
$\phi$ is  necessarily added ``by hand'' to simultaneously implement  this symmetry
and generate the Einstein action. Here $\phi$ is naturally present
because  $\tilde R^2$ is Weyl-covariant.

Assuming that $\phi$ acquires a vev (at quantum level, etc) and after applying a
transformation (\ref{WG}), (\ref{WG2}) with $\Sigma=\phi^2/\langle\phi^2\rangle$,
then Einstein gravity is generated in the broken phase, as an effective theory\footnote{
  One may be concerned about the term $C_{\mu\nu\rho\sigma}^2$ in the action which, as a higher derivative,
  generates a ghost. Since the mass of the ghost is $m\sim \eta M_p$, as long as
  $\eta$ is not fine tuned to values $\ll 1$, the ghost is massive,  can be integrated out 
 and  there is no instability in the theory \cite{Hawking}. In the general
 (non-integrable) Weyl geometry and SMW, one can actually argue that,
 since the metric and connection (equivalently,
 $\w_\mu$) are independent, and treated as such by the variation principle,
 there is actually no ghost \cite{Wheeler}.}
  \bea\label{E}
  \cL_1=\sqrt{g} \,\,
  \Big\{ -\frac12 \,M_p^2 \,R -  \,M_p^2\,\Lambda- \frac{1}{\eta^2}\,C_{\mu\nu\rho\sigma}^2\Big\},
  \qquad \Lambda\equiv \frac14\,\langle\phi\rangle^2,
  \quad
M_p^2\equiv \frac{1}{6\xi^2}\,\langle\phi\rangle^2.
\eea

\medskip
We see that $\phi$ generates both the cosmological constant and the Planck scale,
which are thus related, hence the theory  predicts that $\Lambda\not=0$.
This suggests an ultraviolet (UV) - infrared (IR) connection between the associated
physics at $M_p$ (UV) and at $\Lambda$ (IR), respectively.
The hierarchy $\Lambda\ll M_p^2$ if fixed by one initial classical tuning of
the dimensionless  $\xi\ll 1$ that fixes $M_p$, while the vev of $\phi$ fixes $\Lambda$.
Since we have the equation of motion  $\phi^2=-\tilde R$   (see also Appendix~\ref{BB})
on the ground state $\langle\phi^2\rangle=- R$, therefore  $R=-4\Lambda\not=0$.
Finally, if one considers a Friedmann-Lema\^itre-Robertson-Walker metric one
has that $R=-12 H_0^2$ and  $\Lambda=3 H_0^2$, where $H_0$ is the
Hubble constant.

Note that  $\Lambda$ is positive; in general this is not obvious
in  models with local scale symmetry which are linear in the scalar curvature,
formulated  in either Weyl or Riemannian geometry;
in such cases the coefficient of $\phi^4$ in the action that generates the cosmological constant
is not constrained (as here) and can have any sign/value. We thus see 
the advantage of the Weyl  geometry and its initial {\it quadratic} action (\ref{L1})
that generates a positive $\Lambda$.

To conclude, one can have a local Weyl invariant theory such as 
Weyl's quadratic action that is associated to integrable geometry and
recovers Einstein gravity and predicts a positive $\Lambda$.
Both $M_p$ and $\Lambda$ have ultimately a geometric origin \cite{non-metricity}
because the scalar $\phi$ that generates them is propagated by $\tilde R^2$ which is geometric by nature.
No  scalar field was added to this purpose. This field decouples in the Einstein gauge (frame)
where the symmetry is broken.
In the general (non-integrable) Weyl geometry with gauged Weyl symmetry,
this scalar field is ``eaten'' (in a Stueckelberg mechanism) by the Weyl gauge field
which becomes massive  and subsequently decouples and Einstein gravity
is again recovered \cite{Ghilen1}.
For a further  discussion of local versus gauged Weyl symmetry and
the integrable versus general Weyl geometry see \cite{non-metricity}.
This ends our review of Weyl geometry, its symmetries and associated Lagrangians.

\section{SM in Weyl  geometry: local vs gauged scale symmetry}\label{SM}

Consider now adding the SM to the previous Lagrangian $\cL_1$ in quadratic gravity with
local Weyl symmetry, in other words consider the SM in Weyl  integrable geometry.
This is possible under the  assumption of vanishing Higgs mass, in which case the model
has a local Weyl symmetry. Such embedding is  naturally minimal, without the need for new degrees of
freedom beyond those of the SM and Weyl integrable geometry ($\phi$ and $g_{\mu\nu}$).
This is different from models with this symmetry in Riemannian geometry, as we discuss.
We study the spontaneous breaking of this symmetry and  compare this model to the SMW model \cite{SMW}
obtained by endowing the SM with a gauged Weyl symmetry in general
(non-metric) Weyl conformal  geometry. For convenience, technical details of the SMW 
are reviewed in Appendix~\ref{BBB}.

\subsection{SM with local Weyl symmetry and  integrable geometry}\label{s31}

Consider first the SM  Higgs sector.  The action of the Higgs and gravity  is
  \bea\label{higgsR}
\cL_H\!=\!\sg\,\Big\{\,
\frac{\tilde R^2}{4!\,\,\xi^2}
 -\frac{\tilde C_{\mu\nu\rho\sigma}^2}{\eta^2}
\,-\,\frac{\xi_h}{6}\,\vert H \vert^2 \tilde R +\vert \tilde D_\mu H\,\vert^2
-{\lambda}\, \vert H\vert^4 
\Big\}.
\eea

\medskip\noindent
As mentioned, in Weyl integrable geometry the length curvature tensor which is actually
the field strength of $\w_\mu$ is vanishing $F_{\mu\nu}=0$ hence there is no kinetic term for $\w_\mu$.
This is the only difference in $\cL_H$ and in the total  action  from the case of SMW,
see  Appendix~\ref{BBB} and \cite{SMW}.
This means that locally the  Weyl field is ``pure gauge'', assuming no topological restrictions.
Above we introduced
 \medskip
\bea\label{deriv}
\tilde D_\mu H&=&\big[\partial_\mu - i \cA_\mu - (1/2)\, \q\,\w_\mu\big]\, H,
\\[5pt]
\vert \tilde D_\mu H\vert^2&=&\vert (\partial_\mu - \q/2\,\, \w_\mu) H\vert^2
- i H^\dagger \,(\overleftarrow\partial_\mu \cA^\mu -\cA^\mu \partial_\mu )\,H
+ H^\dagger \cA_\mu \cA^\mu H,
\label{deriv2}
\eea

\medskip\noindent
where\footnote{Also
$
\vert \tilde D_\mu H\vert^2\!
=
\vert D_\mu H\vert^2 -(\q/2) \w^\mu
\big[\nabla_\mu (H^\dagger H)- (\q/2) \,\w_\mu H^\dagger H \big],
$ with $D_\mu$ the SM  covariant derivative.}
 $\cA_\mu=(g/2) \,\vec \tau\cdot \vec A_\mu+ (g^\prime/2)\, B_\mu$
 with  Pauli matrices $\vec\tau$; $\vec A_\mu$,
$B_\mu$ are  the SU(2)$_L$ and U(1)$_Y$ gauge bosons; $\cA_\mu \cA^\mu=
(g/2)^2 \vec A_\mu.\vec A^\mu + (g\,g^\prime/2)\,(\vec \tau.\vec A_\mu)\,B^\mu$
with $g, g^\prime$ the SU(2)$_L$ and U(1)$_Y$ couplings.
Each term in $\cL_H$ is  invariant under (\ref{WG}), (\ref{WG2}).

In  $\cL_H$ we again
replace $\tilde R^2\ra -2\phi^2 \tilde R-\phi^4$ to find a classically equivalent
action; indeed, using the equation of motion of $\phi$ from the new
action and its solution $\phi^2=-\tilde R$
back in the action, one recovers (\ref{higgsR}). This implicitly assumes that $\phi$
is non-vanishing. After this replacement,
the initial higher derivative action is ``unfolded'' to a second order theory;
the non-minimal coupling term in (\ref{higgsR}) is then modified into
\bea\label{rep}
-\frac{1}{6}\,\xi_h\,\vert H\vert^2
\,\tilde R \ra \frac{-1}{12} \,\Big(\,\frac{1}{\xi^2} \phi^2+\,2\,\xi_h\,H^\dagger H\Big)\,\tilde R.
\eea
Then
\be\label{action2}
\cL_H=\sqrt{g}\,\Big\{\frac{-1}{2}\Big[ \frac16 \theta^2 R +(\partial_\mu\theta)^2
-\frac{\q}{2}\, \nabla_\mu (\theta^2 \w^\mu)\Big]
+\frac{\q^2\,\theta^2}{8}\, \Big[\w_\mu
-\frac{1}{\q}\,\nabla_\mu \ln\theta^2\Big]^2
\!\!+\vert \tilde D_\mu H\vert^2
-  V
-\frac{C_{\mu\nu\rho\sigma}^2}{\eta^2}\Big\}
\ee

\medskip\noindent
with $\theta$  the radial direction in the field space of initial $\phi$ and $H$:
\medskip
\bea\label{th}
\theta^2\equiv (1/\xi^2)\,\,\phi^2+ 2\,\xi_h\, H^\dagger H.
\eea

\medskip\noindent
Notice that with perturbative $\xi$ and $1/\xi\gg 1$,  $\theta$ is essentially due to  $\phi$,
for comparable field values of $\phi$ and $H$ and perturbative $\xi_h$.   Also
\medskip
\bea\label{vv}
V=\frac{1}{4!}\,\Big[24\, \lambda\vert H \vert^4
  +\xi^2\, \big(\theta^2-2\,\xi_h\,\vert H\vert^2\big)\Big].
\eea

\medskip
The equation of motion of $\w_\mu$ from $\cL_1$ of (\ref{action2}) has locally a solution
\bea\label{w}
\w_\mu=\frac{1}{\q}\,\nabla_\mu\ln\big(\theta^2+ 2\,H^\dagger H\big).
\eea

\medskip
In the literature, instead of starting with a vanishing length curvature tensor ($F=0$), as we did,
one often assumes that $\w_\mu=(1/\alpha)\partial_\mu \ln \chi^2$ so it is a ``pure gauge'' field
($\omega$ exact one-form, hence it is closed, $F=0$); here
  $\chi$  is some  real scalar field that is found via its equation of motion, 
  leading to the same solution (\ref{w}).

Next, we use solution  (\ref{w})  back in $\cL_H$. We do this below, in
the unitary gauge\footnote{We could still proceed with a general gauge,
  then extra terms would appear in (\ref{action3}), from the rhs of eq.(\ref{deriv2}).
  In addition, in  eq.(\ref{action3})  one would replace
$\sigma/\theta\ra \Arcsinh (\sqrt{2 H^\dagger H}/\theta)$, etc.}
  for the electroweak interaction. Therefore we set
$H=(1/\sqrt{2}) \,h\,\zeta$, where $\zeta^T\equiv (0,1)$ and $2 H^\dagger H=h^2$
where $h$ is the neutral Higgs.
Then
\medskip
\bea
&& 2 H^\dagger \cA_\mu\cA^\mu H = \, h^2\,\cE\, 
\eea

\medskip\noindent
with notation ($W_\mu$ and $Z_\mu$ are the electroweak gauge bosons)
\medskip
\be
\cE\equiv
(g^2/2)\, W_\mu^+ \,W^{- \mu}  + [(g^2+g^{\prime 2})/4]\,Z_\mu Z^\mu.
\ee

\medskip\noindent
For  convenience we also replace original higgs $h$ by  $\sigma$ where
\medskip
\bea\label{h}
h^2=\theta^2\,\sinh^2 \frac{\sigma}{\theta}\qquad \Ra \qquad
\w_\mu=\frac{1}{\alpha}\,\nabla_\mu \ln\Big[\theta^2 \,\cosh^2\frac{\sigma}{\theta}\Big].
\eea

\medskip\noindent
$\sigma$ will be the actual  canonical Higgs field.
Using eqs.(\ref{w}) to (\ref{h}) back in  action (\ref{action2}) then 
\medskip
\bea\label{action3}
\cL_H=\sqrt{g}\,\Big\{
\frac{-1}{2}\Big[\,\frac{\theta^2}{6} R + (\partial_\mu\theta)^2\Big]
+
\frac{\theta^2}{2}\,\partial_\mu \Big(\frac{\sigma}{\theta}\Big) \,
\partial^\mu\Big(\frac{\sigma}{\theta}\Big)
+\frac12\,\cE\, \theta^2 \sinh^2\frac{\sigma}{\theta}\,
-
\,\frac{1}{\eta^2} \,C_{\mu\nu\rho\sigma}^2
-
V\Big\},
\eea

\medskip\noindent
where
\medskip
\bea\label{potential1}
V=\frac{1}{4!}\,\theta^4 \Big\{ 6\lambda\,\sinh^4\frac{\sigma}{\theta}
+
\xi^2\,\Big[1-\xi_h\,\sinh^2 \frac{\sigma}{\theta} \Big]^2
\Big\}.
\eea

\medskip
Since $\theta$ has a non-vanishing vev (this follows from the initial
assumption $\langle\phi\rangle\not=0$, that allowed the linearisation of
$\tilde R^2$ term),  we can now fix the gauge of the Weyl local scale symmetry.
We choose a  gauge where
\medskip
\bea\label{gaugefixing}
\langle\theta^2\rangle=6 \,M_p^2
\eea

\medskip\noindent
where $M_p$ is the Planck scale and $\theta$ is defined in (\ref{th}).
In this gauge the kinetic terms in $\cL_H$ are then canonical and the radial direction
field $\theta$ actually decouples from the spectrum. In the case of SM
with gauged Weyl symmetry embedded in general Weyl conformal symmetry
where $\w_\mu$ is actually dynamical,
the field $\theta$ is ``eaten'' \`a la Stueckelberg by $\w_\mu$ which then
becomes massive \cite{Ghilen1} (see also \cite{SMW}). This is the main difference
from the local Weyl symmetry case  discussed here.

The action  becomes
\smallskip
\bea\label{action4}
\cL_H\!\!\!\!&=&\!\!\!\sqrt{g}\,
\Big\{
\frac{-1}{2}\, M_p^2 R 
+
\frac12\,(\partial_\mu\sigma)^2
+m_W^2(\sigma) W_\mu^+ W^{-\mu} + \frac{m_Z^2(\sigma)}{2}\,Z_\mu Z^\mu
-\frac{1}{\eta^2} C_{\mu\nu\rho\sigma}^2
-V\Big\}\qquad
\eea

\medskip\noindent
Hence Einstein action is obtained  in the broken phase of local Weyl symmetry.
Here we introduced the notation
\bea
m_W^2(\sigma)&\equiv&\frac32\, M_p^2 \,g^2\sinh^2\frac{\sigma}{M_p\sqrt{6}}=
\frac14\, g^2\,\sigma^2 \Big[1+ \frac{\sigma^2}{18 \,M_p^2}+\cdots\Big]  \\
m_Z^2(\sigma)&\equiv &\frac32\,(g^2+g^{\prime 2})\sinh^2\frac{\sigma}{M_p\sqrt{6}}=
\frac14 \,(g^2+g^{\prime^2}) \,\sigma^2\,\Big[1+\frac{\sigma^2}{18 M_p^2}+\cdots\Big]
\eea
for the higgs-dependent ``masses'' of $W^\pm$, $Z$ bosons and finally
\bea\label{potential2}
V&=&\frac{3}{2}\,M_p^4 \,\Big\{ 6\lambda\,\sinh^4\frac{\sigma}{M_p\sqrt 6}
+
\xi^2\,\Big( 1-\xi_h\,\sinh^2 \frac{\sigma}{M_p\sqrt 6} \Big)^2
\Big\}.
\\[6pt]
&=&
\frac14\,\Big[\lambda-\frac19\,\xi_h\,\xi^2
+\frac16\,\xi_h^2\,\xi^2\Big]\sigma^4
-\frac12\,\xi_h\,\xi^2 M_p^2\,\sigma^2
+ \frac{3}{2}\xi^2\,M_p^2
+\cO(\sigma^6/M_p^2).
\eea

\medskip\noindent
An expansion was made for $\sigma\ll M_p$,
to obtain a SM-like Higgs potential. 
The potential in (\ref{potential2}) is
similar to that in SMW \cite{SMW} which has an enlarged, gauged scale symmetry
and  where it  was analysed in detail.  Differences remain however in the Higgs sector,
as discussed shortly. The above expansions show the emergence of higher dimensional
operators corrections in the  EW sector of SM, suppressed by $M_p$.
We also find (see also Appendix~\ref{b3} and \ref{BBB})
\bea\label{Lambda2}
\Lambda
=\frac{1}{4} \langle\phi^2\rangle
=\frac{\xi^2 \langle\theta\rangle^2}{4\,(1+\xi^2\,\xi_h^2/(6\lambda)}
\approx \frac{\xi^2\langle\theta^2\rangle}{4}+\cO(\xi^4\xi_h^2/\lambda).
\eea

\medskip
Note that again  $\Lambda$ and $M_p$ have a common origin being $\propto \langle\theta\rangle$.
A hierarchy $\Lambda\ll M_p$ is then due to the fact that gravity is ultraweak $\xi\ll 1$,
where $\xi$ is the coupling of the $\tilde R^2$ term.

Consider next the SM gauge sector.
The SM  gauge bosons action  is invariant  under transformation (\ref{WG}),
since the gauge bosons do not transform.
One way to see this  is that they enter under the corresponding
covariant derivative acting on a field charged under it and should transform
(have same weight) as  $\partial_\mu$ acting on that field;
since coordinates are kept fixed under (\ref{WG}), the gauge fields do not transform
under (\ref{WG}) i.e. have a vanishing Weyl charge.
Then their kinetic terms are similar
to those in the SM in pseudo-Riemannian geometry since the Weyl connection is symmetric:
 $(F_{\mu\nu})_{ SM}$ involves the difference $\tilde\nabla_\mu A_\nu -\tilde\nabla_\nu A_\mu$, where
$A_\mu$ denotes a SM gauge boson and
since $\tilde\nabla_\mu A_\nu=\partial_\mu A_\nu -\tilde\Gamma_{\mu\nu}^\rho A_\rho$ for a symmetric
$\tilde\Gamma_{\mu\nu}^\rho=\tilde\Gamma_{\nu\mu}^\rho$ then $\tilde\Gamma_{\mu\nu}^\rho$  
(and $\w_\mu$)  cancels out in $F_{\mu\nu}$ which then  has a form as in flat space-time.
Thus the gauge kinetic term does not depend on the Weyl connection and is similar
to that  in pseudo-Riemannian geometry:
\bea
\cL_g=-\sum_{\rm{groups}}\frac{\sg}{4} g^{\mu\rho} g^{\nu\sigma} F_{\mu\nu} F_{\rho\sigma},
\eea
where the sum is over the SM gauge group factors:  $SU(3)\times SU(2)_L\times U(1)_Y$.

Finally, consider the SM fermions sector.
According to (\ref{WG2}) the  fermions have non-zero Weyl weight. 
However, the fermions action is Weyl vector field independent \cite{Kugo} (for a review \cite{SMW})
and is invariant under transformation (\ref{WG}), (\ref{WG2}). 
This invariance  is due to a cancellation in the Dirac action
of the dependences on $\w_\mu$ between that of
the Weyl-covariant derivative acting on fermions and that of the spin connection part of this derivative.
Since the fermions action is independent of $\w_\mu$ in Weyl geometry,
this remains true in our particular integrable  case of $\omega_\mu$ of (\ref{w}). So the fermions
action is similar to the (pseudo)Riemannian case
\medskip
\bea\label{Lf}
\cL_f & =&\frac12\, \sg\,\, \overline\psi \, i\,\gamma^a \, e^\mu_{\,a}\,
\,\Big[
\partial_\mu - i g\, \vec T \vec{\hat A}_\mu - i\, Y g^\prime \hat B_\mu
+\frac12 s_\mu^{\,\,ab} \,\sigma_{ab} \Big]\,\psi+h.c.,
\eea

\medskip\noindent
with the usual quantum numbers of fermions under SM group and with  $\vec T=\vec\tau/2$.
The spin connection has the usual form of Riemannian geometry
\bea
s_\mu^{ab}=-e^{\lambda\,b}\,(\partial_\mu\,e^a_\lambda -\Gamma_{\mu\lambda}^\nu\, e^a_\nu).
\eea

One can check that $\cL_f$ is invariant under (\ref{WG}). 
Regarding the Yukawa interaction $\cL_Y$ it
 is similar to that in flat space-time uplifted to curved space-time and
 is also invariant under (\ref{WG}), (\ref{WG2}), as seen from Weyl charges of the fields involved
 and of $\sqrt{g}$ \cite{SMW}.
 
 The SM Lagrangian in Weyl integrable geometry is then given by 
 \medskip
 \bea\label{action}
\cL= \cL_H+\cL_f+\cL_g+\cL_Y.
\eea

\medskip\noindent
  The only difference between our case
 and SMW \cite{SMW} is thus manifest in the Higgs sector $\cL_H$.
 
In conclusion,  the SM action changes very little when  this is
endowed with  local scale symmetry in quadratic gravity in the context of Weyl integrable geometry.
No new degrees of freedom beyond those of the SM and underlying geometry were added,
the Einstein action is easily recovered in the broken phase while in the EW sector
the higher dimensional operators generated are suppressed by $M_p$. Thus, the SM admits a
truly minimal and natural embedding in Weyl integrable geometry, which is interesting.

\subsection{Inflation predictions}

Here we comment on an immediate phenomenological prediction.
The model can have successful inflation based on the potential in eq.(\ref{potential2}).
The potential is similar to that  discussed in  \cite{WI3,WI1} see also \cite{WI2,SMW,Ghilen1,Palatini},
for which inflation predictions were already studied in detail, hence we can use below these results.

Although the potential depends on the Higgs field only, the inflation mechanism is of
Starobinsky-Higgs type.
This is understood as follows: successful inflation
requires that the second term in eq.(\ref{potential2}) dominates and gives
the leading (flat) contribution during inflation. This contribution is actually that
due to the field  $\phi$, as seen by comparing eqs.(\ref{th}) and (\ref{vv}),
with $\phi$ due to the  $\tilde R^2$ term in the action.
This  explains the expectation for  inflation predictions similar to
those in  Starobinsky model. 
One can show that the potential in (\ref{potential2}) gives a dependence
of the tensor-to-scalar ratio ($r$) on the spectral index $n_s$ \cite{WI1,WI2}
which is
\bea
r=2( 1-n_s)^2-\frac{16}{3}\,\xi_h^2 +\cO(\xi_h^3).
\eea
%
The  term dependent on $\xi_h$ is due to the non-minimal coupling of the
Higgs field and its effect can be ignored  for
$\xi_h<10^{-3}$ since then its correction to $r$ is too small to be measured.
In such case the above relation is similar to that in 
Starobinsky inflation,  where $r=3(1-n_s)^2$.  The value of $r$ in Starobinsky inflation
is an upper bound  that  is saturated in the present model for $\xi_h\ra 0$.
For larger $\xi_h\sim 10^{-3} - 10^{-2}$, a smaller $r$ is obtained. Numerically,
for $n_s=0.9670\pm 0.0037$ at 68$\%$ CL then \cite{WI1,WI2,WI3} (see also \cite{Tang})
\medskip
\bea
0.00257\leq r \leq 0.00303, 
\eea
This result  demands  $\lambda\ll \xi_h^2 \,\xi^2$ which can be
respected for small enough  $\lambda$. Such value for $r$ is within the 
reach of  the new  generation of CMB experiments:
CMB-S4, LiteBIRD, PICO, PIXIE  \cite{CMB1,CMB4,CMB3,litebird,Pixie,CMB2}.

\subsection{Comparison to SMW}

Let us now compare our  model to the  SM with gauged scale symmetry \cite{SMW} (SMW) in general Weyl geometry.
The details of the SMW action and its equations of motion  are presented in
Appendix~\ref{BBB}, for convenience.  The difference between the action of the SM with local
Weyl symmetry and the SMW is manifest   in the Higgs and gravity sectors ($\cL_H$) and is given by
the absence of $F_{\mu\nu}^2$  in the initial Lagrangian eq.(\ref{higgsR}) of the current model.
Let us then compare the broken and symmetric phases of the final actions of the two theories.

In the broken phase, the Einstein action is naturally recovered in both cases; the difference is that
in the SMW action there are additional terms $\Delta\cL$ present \cite{SMW},
compared to action  (\ref{action}) of the case with local Weyl symmetry, with
\bea
\Delta \cL&=&\sqrt{g}\,\,
\frac{3}{4}\,M_p^2\,\q^2\,\w_\mu \,\w^\mu \Big[1+
\sinh^2 \frac{\sigma}{M_p\sqrt{6}}\Big]-\sqrt{g}\,\,\frac14 \,F_{\mu\nu} F^{\mu\nu}
\\
&=&\sqrt{g}\,\,\Big\{
\frac34 \,\q^2\,M_p^2 \,\w_\mu \w^\mu
+ \frac18\,\q^2 \,\sigma^2 \,\w_\mu\,\w^\mu
+ \cO(1/M_p^2)\Big\}
-\sqrt{g}\,\,\frac14 \,F_{\mu\nu} F^{\mu\nu}
\eea
where $F_{\mu\nu}$ is the field strength of $\w_\mu$. 
The mass  term for $\w_\mu$ follows from a Stueckelberg mechanism in which $\w_\mu$ has
eaten the $\theta$ field to become massive \cite{Ghilen1}.
$\Delta \cL$ contains an important new Higgs coupling $\sigma^2\,\w_\mu \w^\mu$ absent in our
current model. This coupling comes from the Higgs kinetic term $\vert\tilde D_\mu H\vert^2$.
These are the  main differences between the current model and SMW.
In SMW  there can also be a gauge  kinetic mixing  term
between $\w_\mu$ and the SM $U(1)_Y$  field, not shown here; however, such mixing  is strongly constrained
 from its correction to the  $Z$ boson mass \cite{SMW}.

We see that the {\it broken phase} of  local Weyl symmetry  of our model
is indeed recovered from that of SMW \cite{SMW} by simply decoupling
the massive Weyl vector field in the action. For a large enough mass $m_\w$ of $\w_\mu$,
the correction $\Delta \cL$ is  suppressed and one is left with the action of the
present case.  Naively, one would think that $m_\w$ is very large (near $M_p$)
but in fact $m_\w$ can be significantly lower. Actually, the
current lower bound on $m_\w$ which  sets the so-called
``non-metricity scale'' is of few TeV only \cite{Latorre}.
This  bound is found by demanding that non-metricity  corrections (i.e. due to $\w_\mu$) to the
Bhabha scattering cross section be within the current  error of this cross section;
this bound is found by using a Dirac action that involves
non-metricity ($\w_\mu$)  \cite{Latorre} which mediates this process;
 it must be said, however,  that in Weyl geometry there is actually no coupling of SM
fermions to vectorial non-metricity ($\w_\mu$) \cite{Kugo}
(see also \cite{SMW}), hence this non-metricity lower  bound  can actually be evaded and be
 lower than few TeV. This deserves further study.

Regarding the symmetric phases,  the local Weyl symmetry
of  the current model  is naturally enlarged to a Weyl gauge symmetry in SMW
\cite{SMW} that brings in the $\w_\mu$ gauge field, see eqs.(\ref{WG}), (\ref{WG2}), (\ref{WGg}).
The SMW with gauged Weyl symmetry seems then  a more general and physical UV completion
of the SM and Einstein gravity into a {\it gauge theory} of scale invariance
than the current model with local Weyl symmetry only (no $\w_\mu$). To
understand why this is so, note that in  the SMW  there is an associated
conserved current \cite{Ghilen1}, see eq.(\ref{ccc}) 
\bea\label{J}
\nabla^\mu J_\mu=0,\qquad
J_\mu=\frac{\alpha}{4}\,\big(\nabla_\mu - \alpha\,\w_\mu\big)\, K,
\qquad K\equiv \theta^2+h^2.
\eea

This relation  is a generalisation 
of an onshell current  conservation in {\it global} scale invariant
theories  \cite{Ga,Fe1,Fe2,Fe3}.
In our current model, however, $\w_\mu$ is that of  eq.(\ref{w}), which when used in (\ref{J})
gives $J_\mu=0$. This means that the current associated with a local Weyl
symmetry is trivial and the charge is vanishing if $\w_\mu$ is not dynamical.
This seems a general problem with models with local Weyl symmetry \cite{J1,J2},
regardless of the underlying geometry considered (Weyl or Riemannian) where this symmetry is implemented.
This  questions if local Weyl symmetry is physical and
if it can be a symmetry of a fundamental, UV-complete theory.

In the light of the above arguments, we conclude
that SMW is a more fundamental UV completion of SM and Einstein gravity
than its version with local Weyl symmetry discussed here. SMW provides 
such UV completion in a full gauge theory of scale invariance\footnote{It
  remains to be seen if it is also renormalizable as in quadratic gravity
\cite{Stelle}, see \cite{SMW,non-metricity}
for some arguments.}. Correspondingly, the (non-metric) Weyl conformal geometry underlying the SMW
is more fundamental than its (metric) integrable version underlying the current model,
that is conformal to Riemannian geometry of Einstein gravity.
And like  any gauge symmetry, the  gauged
Weyl symmetry of SMW must remain valid at quantum level. 
Loop corrections thus require a regularisation/renormalization
that respects this symmetry  \cite{Englert} (also more recent \cite{Misha,Ghilen2}).
In this way the symmetry is maintained at quantum level and one
 avoids the  Weyl anomaly, see \cite{WG} for an update.
The UV completion of SM and Einstein gravity  can thus   be realised by
a (quantum) gauged Weyl invariant theory. This deserves further study.

\section{Conclusions}

Since the SM with the Higgs mass parameter set to zero has  global scale symmetry,
we explored the consequence of making this symmetry local and including gravity,
using the gauge invariance principle.
We studied the SM in  quadratic gravity with {\it local} Weyl symmetry which is
of strong interest in gravity theories, and compared it to SM in quadratic gravity with a
{\it gauged} Weyl symmetry (SMW) studied previously; these models are distinguished by the
absence or presence of an associated Weyl gauge boson of dilatations ($\w_\mu$), respectively.

We  showed that  SM and Einstein gravity admit a natural embedding in
so-called Weyl  integrable geometry, with no new fields needed
beyond the SM and this geometry. This geometry is a special limit of Weyl conformal  geometry
of vanishing length curvature  tensor ($F_{\mu\nu}=0$).
Physically, this means that $\w_\mu=0$  or is  ``pure gauge'' (non-dynamical),
as in  conformal gravity. Hence, the  theory  has   local Weyl invariance, but there is no gauge boson.
The theory is  quadratic in the scalar curvature,
so is more general than  previous theories of SM with local Weyl symmetry
in (pseudo)Riemannian geometry which were linear in scalar curvature; these
required a scalar  field $\phi$ (beyond SM) be added ``by hand''
to implement this symmetry and  generate Einstein gravity from $\phi^2 R$.
The situation also  differs from the SMW which has the  larger, {\it gauged}
Weyl symmetry in general Weyl geometry, where $\w_\mu$ is dynamical, but then
the geometry becomes non-metric.

The local Weyl symmetry of our theory is spontaneously broken
to Einstein action. The mass scales (Planck scale 
and cosmological constant $\Lambda$)  have geometric origin,
being generated by a scalar field $\phi$ arising from the $\tilde R^2$ term in the action.
$\Lambda$ is positive due to  the  quadratic  nature of the action. This situation is
similar to the SMW where,  in addition, $\ln\phi$ is further  ``eaten''
by the Weyl gauge boson $\w_\mu$ which then becomes a massive Proca field (by Stueckelberg mechanism)
and subsequently decouples. Therefore, after decoupling,  the broken phase of SMW recovers
the broken phase of local Weyl symmetry of our current model, as we verified.

The hierarchy of the scales $\Lambda\ll M_p$  is generated by one classical tuning of the
 dimensionless coupling $\xi$ of the $\tilde R^2$ term in the action. One may expect
this hierarchy remain stable at a quantum level, due to local  Weyl symmetry.
Inflation is Starobinsky-like, with the inflaton role
essentially played by the same $\phi$ coming from the  $\tilde R^2$ term, as in SMW,
and  the scalar potential is similar in both theories. 
The tensor-to-scalar ratio $r$ is then  similar to that in SMW,
$r \sim 3 \times 10^{-3}$, with an upper bound equal to
the Starobinsky model  value, saturated for a vanishing Higgs non-minimal coupling.
The present theory differs however  from the SMW in the Higgs sector: SMW  predicts
a direct coupling of the Higgs $\sigma$ to the Weyl gauge boson $\q^2\,\omega_\mu\omega^\mu \sigma^2$,
the implications of which are yet to be explored.

There is one issue that seems common to all models beyond SM and Einstein gravity
with local Weyl symmetry, regardless of their underlying geometry:
the associated current of this symmetry vanishes. This raises concerns on the
physical meaning of such symmetry. 
The gauge invariance principle tells us that
one should actually implement  the full gauged Weyl symmetry. The SMW
with its (larger) gauged Weyl  symmetry and (non-trivial) conserved current
is therefore more fundamental. In conclusion, SMW
is a good UV completion  of  the current theory and of SM,
giving an (anomaly-free) {\it gauge theory} of scale invariance  that
generates Einstein gravity in its broken phase, as an effective theory.

\bigskip
\bigskip
 \section*{Appendix}

\def\theequation{A-\arabic{equation}}
\def\thesubsection{A}
\setcounter{equation}{0}
\def\thefigure{A-\arabic{figure}}
\def\thelabel{A}
 
 \subsection{Weyl geometry and parallel transport}\label{AppendixA}

 We review here the  parallel transport in Weyl geometry and its integrable limit.
 As stated in the text,
Weyl geometry is represented by  classes of equivalence of $(g_{\mu\nu}, \w_\mu)$ related
by (\ref{a3}). If present, scalars $\phi$ and fermions $\psi$ also transform
\bea\label{a3}
&&\hat g_{\mu\nu}=\Sigma^q \,g_{\mu\nu}, \quad
\sqrt{\hat g}=\Sigma^{2\,q}\sqrt{g},\quad
\hat\w_\mu=\w_\mu-\frac{1}{\alpha}\,\partial_\mu\ln\Sigma,\quad
\hat e^a_\mu=\Sigma^{q/2}\, e^a_\mu,\quad
\\[5pt]
&&\,\,\,\,\hat\phi=\Sigma^{-q/2}\,\phi,\quad
\hat\psi=\Sigma^{-3 q/4} \psi.\quad
\label{a4}
\eea
Here $q$ is the Weyl charge of the metric (in the text we set $q=1$).
Weyl geometry is non-metric which means $\tilde\nabla_\mu g_{\alpha\beta}\not=0$, or more exactly:
\bea\label{wgeom}
\tilde\nabla_\mu g_{\alpha\beta}=-\alpha\,q\, \w_\mu \,g_{\alpha\beta},
\eea
where $\tilde\nabla$ is defined by the Weyl connection $\tilde\Gamma$
\bea
\tilde\nabla_\mu g_{\alpha\beta}=\partial_\mu g_{\alpha\beta}
-\tilde\Gamma_{\alpha\mu}^\rho g_{\rho\beta}
-\tilde\Gamma_{\beta\mu}^\rho g_{\rho\alpha}.
\eea
Eq.(\ref{wgeom})  may be written as
\bea
\tilde\nabla^\prime _\mu g_{\alpha\beta}=0,
\qquad \tilde\nabla^\prime\equiv \tilde\nabla\Big\vert_{\partial_\mu\ra\partial_\mu +
  \alpha\, q\, \w_\mu}.
\eea
Therefore the (symmetric)  Weyl connection $\tilde\Gamma$ is found from the
Levi-Civita connection ($\Gamma$) in which
one makes the same substitution: $\tilde\Gamma=\Gamma\vert_{\partial_\lambda\ra\partial_\lambda +
  \alpha\, q\, \w_\lambda}$. This gives
\bea\label{tGammas}
\tilde \Gamma_{\mu\nu}^\lambda=
\Gamma_{\mu\nu}^\lambda+(q/2)\,\q \,\Big[\delta_\mu^\lambda\,\, \w_\nu +\delta_\nu^\lambda\,\, \w_\mu
- g_{\mu\nu} \,\w^\lambda\Big].
\eea
Consider now  an arbitrary  vector $u^\mu$ of some non-vanishing Weyl charge ($z_u\not=0$):
\bea
 \hat u^\mu =\Sigma^{z_u/2} u^\mu
\eea
The parallel transport of a constant vector  (in a Weyl-covariant sense) is defined by
\be\label{pt}
\frac{D\, u^\mu}{d\tau}=0, \quad \text{where}\quad
D\equiv dx^\lambda\,D_\lambda,\quad
D_\lambda\, u^\mu= \tilde\nabla_\lambda u^\mu \Big\vert_{
 \partial_\lambda\ra \partial_\lambda+(z_u/2)\,\alpha\,\w_\lambda},
\ee
with 
\bea
\tilde\nabla_\lambda u^\mu=
\partial_\lambda u^\mu
+\tilde\Gamma_{\lambda\rho}^\mu\,u^\rho,
\eea
and $x=x(\tau)$. Then from (\ref{pt}) the  differential variation of the vector is
\bea\label{A10}
d\, u^\mu=dx^\lambda \,\partial_\lambda u^\mu=
-d x^\lambda \,\Big[ (z_u/2)\, \alpha\,\w_\lambda \,u^\mu +\tilde\Gamma_{\lambda\rho}^\mu\,u^\rho\Big],
\eea
Under  the  parallel transport along a curve, 
a product $\langle u,v\rangle=u^\mu\,v^\nu \, g_{\mu\nu}$ of vectors $u, v$ changes, too, so using
(\ref{wgeom}) and (\ref{A10}) applied to $u, v$ vectors,  one finds
\bea\label{prod}
d\langle u,v\rangle\!\!\!& =& d\,\big[ u^\mu \,v^\nu\,g_{\mu\nu}\big]
=
dx^\lambda \,\big[\tilde\nabla_\lambda g_{\mu\nu}- \alpha\,\w_\lambda\, g_{\mu\nu} (z_u+z_v)/2\big]
u^\mu u^\nu.\\[4pt]
\!\! \! &=&
-\q\,dx^\lambda\,\w_\lambda\, \big[ q+ (z_u+z_v)/2\big]\, \langle u, v\rangle\qquad\quad
\eea
For the norm of a vector 
\bea\label{sss}
d \ln \vert u\vert^2= dx^\lambda \,\w_\lambda\,(-\alpha)\,(q+z_u),
\eea
or, integrating this along a path $\gamma(\tau)$:
\medskip
\bea\label{sym}
\vert u\vert^2=\vert u_0\vert^2\,e^{-\alpha\, (q+z_u)\,\int_{\gamma(\tau)}  \w_\lambda dx^\lambda}.
\eea
 $u$ and $u_0$ are the values at the end points of the path.
 Hence, using (\ref{a3}) we see that,  in general,  the integral and
 the norm  of the vectors are path-dependent in Weyl geometry
 (in Riemannian case $\omega_\lambda=0$ and the norm is invariant
 for any $z_u$ and any $\gamma$).
 
However, for any  tangent space vector  $u^a=e^a_\mu u^\mu$ that is invariant
i.e. it  has a vanishing charge,  then  $z_u/2+q/2=0$ (since $e^a_\mu$ has charge $q/2$), and with this
the norm of $u^\mu$ itself in eq.(\ref{sym}) is actually invariant $\vert u\vert=\vert u_0\vert$
under parallel transport for any $\gamma$.

Finally,  the ratio of the norms of two  vectors of same but  arbitrary Weyl weight
is also invariant under the parallel transport, for same $\gamma$.  This is seen by using (\ref{sss})
\bea
d\,\ln \frac{\vert u\vert^2}{\vert v\vert^2}=(-\alpha)\,
(z_u -z_v)\,\w_\lambda\,dx^\lambda,
\eea
which  vanishes if $z_u=z_v$: the relative length is then invariant.

In Weyl integrable geometry   $\w_\lambda=\partial_\lambda (\text{scalar-field})$ and
if the path is closed then the integral vanishes and the norm is invariant
$\vert u\vert=\vert u_0\vert$, (and path-independent between 2 points).   This result was used in the past
to favour gravity theories  based on  Weyl integrable geometry (instead of general non-metric Weyl geometry)
since it meant that there was  no second clock effect: the theory is metric, then 
rods' length and clocks' rates would not change.  This  was to avoid long-held
criticisms \cite{Weyl1} directed at  gravity  theories based on  (non-metric) Weyl conformal geometry.

Such  criticisms were misleading for (non-metric) Weyl geometry
since they implicitly  assumed that $\w_\lambda$ was  a  massless gauge field.
Actually, in the symmetric phase there is no second clock effect since one cannot define a clock
rate in the absence of a mass scale (forbidden by the symmetry of the action).
To  test the second clock effect and compare to an experiment in a gauge theory
of scale invariance built in a general  Weyl geometry, one must first
{\it fix the gauge}  of this symmetry.  The gauge  is naturally fixed since 
$\phi$ (which is part of $\tilde R^2$  geometric term in the action)  and has a non-zero  vev, is
``eaten'' in a  Stueckelberg mechanism by  $\w_\lambda$ which thus becomes massive and
decouples \cite{Ghilen1}, and the symmetry is spontaneously broken. In this broken phase
non-metricity effects due to $\w_\mu$ are strongly suppressed by a large enough mass of $\w_\mu$
(the current lower bound on $m_\w$ is only  few TeV \cite{Latorre}!) and the
mentioned criticisms are avoided.
This breaking is  geometric in nature \cite{Ghilen1} and needs no  scalar fields be added
``by hand'' to this purpose. Hence the absence of second clock effect is  general, regardless of the matter
content of the theory embedded in   Weyl  geometry.

\def\theequation{B-\arabic{equation}}
\def\thesubsection{B}
\setcounter{equation}{0}
\def\thefigure{B-\arabic{figure}}
\def\thelabel{B}

\subsection{SM in Weyl geometry and its integrable limit}\label{B1}

We discuss the equations of motion of Weyl quadratic gravity first in the absence, then in the
presence of the SM, for the case with gauged Weyl symmetry then take the limit of
local scale symmetry (integrable geometry case).

\subsubsection{Weyl quadratic gravity}\label{BB}

The review here is  based on \cite{SMW} and gives
the equations of motion for an action with gauged Weyl symmetry, which is similar to
action (\ref{oo}), (\ref{ooo})  (without the $C_{\mu\nu\rho\sigma}^2$ term),
but has in addition a kinetic term for $\w_\mu$.
We then examine what happens if this term is absent (integrable case, $\w_\mu$ ``pure gauge'').
Consider then
\bea\label{altss}
\cL &= &
\sqrt{g}\, \Big\{\frac{1}{4!} \,\frac{1}{\xi^2}\,\tilde R^2-\frac14\,F_{\mu\nu}^2\Big\}
\\
&=&
\sqrt{g}\,\Big\{- \frac{1}{12}\,\frac{\phi^2}{\xi^2}
\,\Big[ R- 3 \q \nabla_\mu\w^\mu -\frac{3}{2}
\q^2\,\w_\mu\w^\mu\Big]
-\frac{1}{4!}\frac{\phi^4}{\xi^2} - \frac14 \,F_{\mu\nu}^2
\Big\}.\label{act2}
\eea
Eq.(\ref{tildeR}) was used and $\tilde R^2$ was linearised with the aid of the
newly introduced field $\phi$, as in the text. After Stueckelberg mechanism \cite{Ghilen1,SMW}
\medskip
\bea
\cL=\sqrt{g}\,\Big\{ -\frac12 \, M_p^2\, R -\Lambda \, M_p^2 -
\frac14 \,F_{\mu\nu}^2 + \frac34\,\q^2\,M_p^2\,\omega_\mu \omega^\mu\Big\},
\quad\Lambda\equiv
\frac{\langle\phi\rangle^2}{4},\quad M_p^2\equiv\frac{\langle\phi\rangle^2}{6\xi^2}.
  \eea
Using the notation:
\bea
\cK=\frac{\phi^2}{\xi^2}, \qquad \cV=\frac{1}{4!} \,\frac{\phi^4}{\xi^2},
\eea
the variation of (\ref{act2})  with respect to the metric gives
\medskip
\bea\label{eqg2ss}
\frac{1}{\sqrt{g}}
\frac{\delta \cL}{\delta g^{\mu\nu}}
&=&\frac{1}{12}\Big\{
-  \cK \Big[ R_{\mu\nu}- \frac12\,g_{\mu\nu} \,R\Big]
- \Big[ g_{\mu\nu}\Box-\nabla_\mu \nabla_\nu\Big] \cK
\nonumber\\
&+ &\frac{3\q^2}{2} \cK \Big[\w_\mu\,\w_\nu- \frac12 \, g_{\mu\nu}\,\w^\rho\,\w_\rho\Big]
-\frac{3\,\q}{2} \Big[ \w_\nu\nabla_\mu+\w_\mu\nabla_\nu - g_{\mu\nu}\,\w^\rho\nabla_\rho\Big]\,\cK \Big\}
\nonumber\\[5pt]
&-&\frac12\,
\Big[
g^{\alpha\beta}\,F_{\mu\alpha} F_{\nu\beta} - \frac14 \, g_{\mu\nu}\,F_{\alpha\beta}\,F^{\alpha\beta}\Big]
+\frac12 g^{\mu\nu} \cV=0.
\eea
Taking the trace
\bea\label{trace1ss}
\frac{1}{12}\,\Big(\cK\, R+ 3 \alpha \,\omega^\mu\,\nabla_\mu \cK
-\frac{3\alpha^2}{2} \, \cK \omega_\mu\omega^\mu-3\Box\cK\Big) +2 \cV=0.
\eea

\medskip\noindent
Further, the equation of motion of $\phi$ that is obtained from the above Lagrangian is
\bea\label{phieqss}
\frac{1}{12\, \xi^2}\, \phi^2 \Big(\tilde R +\phi^2\Big)=0
\eea
This is actually known  ($\phi^2=-\tilde R$), since it was introduced to linearise
the $\tilde R^2$ term.
On the ground state $\langle\phi\rangle^2=-R$ or $R=-4\Lambda$,  which
is also found  from the above trace equation.

For a Friedmann-Robertson-Walker metric one has
$R=-12 H_0^2$ and equation $R=-4\Lambda$ then gives that in Weyl quadratic gravity
\bea
\Lambda=3 H_0^2. 
\eea

\noindent
The equation of motion of $\omega_\mu$ can be written as 
\bea\label{Jmuss}
J^\mu+\nabla_\rho F^{\rho\mu}=0,\qquad\text{where}\qquad
J^\mu\equiv -\frac{\q}{4} g^{\mu\nu}\big(\nabla_\nu - \alpha\omega_\nu)\, \cK
\eea
Using the anti-symmetry in indices of the field strength $F$,
by applying $\nabla_\mu$ on (\ref{Jmuss}) then
\bea\label{cccss}
\nabla_\mu J^\mu=0.
\eea

\medskip\noindent
Therefore, for  a dynamical $\omega_\mu$ we have a non-trivial conserved current $J_\mu$.
This  is also seen by subtracting (\ref{phieqss}) from (\ref{trace1ss}):
one obtains again eq.(\ref{cccss}) with $J^\mu$ as in (\ref{Jmuss}).

Finally, consider the integrable geometry case, when $\omega_\mu$ is non-dynamical in (\ref{altss}),
($F_{\mu\nu}=0$), then the equation of motion of $\omega_\mu$ is
\be
\omega_\mu=\frac{1}{\alpha}\partial_\mu\ln \cK.
\ee
Using this back in  $J^\mu$ one see that $J^\mu$ vanishes. Thus the
associated current to the symmetry is trivial and the symmetry is then ``fake'' \cite{J1,J2},
raising doubts on its physical meaning, in this case.
Hence, in the absence of matter, the current is non-trivial only for a
dynamical Weyl field, showing the importance of fully gauging the scale symmetry.

 \subsubsection{SM in Weyl geometry (SMW)}\label{BBB}

 Here we give details of the SM in Weyl conformal geometry (SMW);
 this information is  complementary to that in \cite{Ghilen1,SMW} and shows
 the Lagrangian, the equations of motion and conserved current.
 We then take the limit  of integrable geometry studied in the text.

 As shown in Section 2 of \cite{SMW}, the SMW action for fermions and gauge bosons  is similar to that
 of SM in  Riemannian geometry, but
 it differs significantly in the Higgs and gravity sectors.
 A gauge kinetic mixing  SM hypercharge - $\omega_\mu$  may also be present in SMW. 
 Hence,  the part of the SMW Lagrangian  of Higgs + gravity + hypercharge sectors
  is \cite{SMW}
 \medskip
  \be\label{higgsRp}
\cL_H\!=\!\sg\,\Big\{\,
\frac{\tilde R^2}{4!\,\,\xi^2}
\,-\,\frac{\xi_h}{6}\,\vert H \vert^2 \tilde R +\vert \tilde D_\mu H\,\vert^2
-{\lambda}\, \vert H\vert^4 
 -\frac{1}{4}  \Big( F_{\mu\nu}^{\,2}+ 2\sin \chi\, F_{\mu\nu}\, F_Y^{\mu\nu} + F_{Y\,\mu\nu}^{\,2}\Big)
-\frac{1}{\eta^2}\tilde C_{\mu\nu\rho\sigma}^2\Big\}.
\ee
which is invariant under (\ref{WG}), (\ref{WG2}), (\ref{WGg})
and with the Weyl-covariant derivative
\bea
\tilde D_\mu H&=&\big[\partial_\mu - i \cA_\mu - (1/2)\, \q\,\w_\mu\big]\, H,
\\[6pt]
 \vert \tilde D_\mu H\vert^2&=&
\vert D_\mu H\vert^2 -\frac{\q}{2}\,\w^\mu\,
\Big[\nabla_\mu (H^\dagger H)- \frac{\q}{2} \,\w_\mu H^\dagger H \Big].
 \eea

 \medskip\noindent
 where $D_\mu H$ is the SM derivative of the Higgs. $F_{\mu\nu}$  ($F_{Y\, \mu\nu}$)
 is the Weyl vector (hypercharge) field  strength, respectively. As in the text, we linearise
 the $\tilde R^2$ term with the aid of a scalar field $\phi$ by replacing in $\cL_H$:
 $\tilde R^2\ra -2\phi^2 \tilde R-\phi^4$. The equation  of motion  of $\phi$ has
 solution $\phi^2=-\tilde R$ which when used back in $\cL_H$  recovers (\ref{higgsRp}).
 Up to a total derivative $\cL_H$ becomes  
\smallskip
\be\label{L2}
\cL_H= \sqrt{g}
\Big\{ -\frac{1}{12} \theta^2\,R -\frac{\alpha}{4} \omega^\mu \nabla_\mu K
+ \frac{\alpha^2}{8} K\,\omega_\mu \omega^\mu
+ \vert D_\mu H\vert^2 - V -\frac{1}{4}\cF_{\mu\nu}^2
-\frac{1}{\eta^2} C_{\mu\nu\rho\sigma}^2\Big\}
\ee

\medskip\noindent
where $\theta$ is the radial direction in the field space of initial fields ($\phi$, $H$), and
\bea\label{L3}
\theta^2&\equiv& \frac{1}{\xi^2}\phi^2+ 2\xi_h\,H^\dagger H,\qquad  K\equiv \theta^2+ 2 H^\dagger H.
\\[2pt]
V&=&\lambda\vert H\vert^4+\frac{1}{24\xi^2}\,\phi^4
=\lambda \vert H\vert^4 + \frac{\xi^2}{24}\,(\theta^2-2\xi_h\,
H^\dagger H)^2.
\\
 \cF_{\mu\nu}^2&\equiv&
\frac{1}{\gamma^2}  F_{\mu\nu}^{\,2}+ 2\sin \chi\, F_{\mu\nu}\, F_Y^{\mu\nu}
+ F_{Y\,\mu\nu}^{\,2},\qquad \qquad\frac{1}{\gamma^2}\equiv 1+\frac{6\alpha^2}{\eta^2}>1.
\label{hhh}
\eea

\medskip
Below we consider the equations of motion in a
unitary gauge $H=1/(\sqrt 2)\, h\, \zeta^T$, $\zeta=(0,1)$.
We shall also ignore the Weyl term $\tilde C_{\mu\nu\rho\sigma}^2$ and hence $\gamma=1$;
this term would not contribute to the trace equation below; this term was studied
extensively in \cite{Mannheim}.
In the unitary gauge
\be
\cL_H= \sqrt{g}\,
\Big\{ -\frac{1}{12} \theta^2\,R -\frac{\alpha}{4} \omega^\mu \nabla_\mu K
+ \frac{\alpha^2}{8} K\,\omega_\mu \omega^\mu
+\frac12 (\partial_\mu h)^2+ \frac12 \,h^2 \,\cE_{\mu\nu} \, g^{\mu\nu}  - V -\frac{1}{4}\,\cF_{\mu\nu}^2
\Big\}\quad
\ee
where
\bea\label{b20}
\theta^2&\equiv& \frac{1}{\xi^2}\phi^2+ \xi_h\,h^2,\qquad  K\equiv \theta^2+ h^2
\\
V&=&\frac14 \lambda\, h^4+ \frac{\xi^2}{24}\,(\theta^2-\xi_h\,h^2)^2.\label{B21}
\\[5pt]
 \cF_{\mu\nu}^2&\equiv&
 F_{\mu\nu}^{\,2}+ 2\sin \chi\, F_{\mu\nu}\, F_Y^{\mu\nu}
+ F_{Y\,\mu\nu}^{\,2},
\\[5pt]
\cE_{\mu\nu}&\equiv&  \frac{g^2}{2}\,W_\mu^+ \,W_\nu^- +\frac{g^2+g^{\prime 2}}{4}\,Z_\mu Z_\nu. 
\eea

\medskip\noindent
The action is Weyl gauge invariant. The equation of motion for $g^{\mu\nu}$ from $\cL_H$ gives
\medskip
\bea\label{eqg}
\frac{1}{\sqrt{g}}
\frac{\delta \cL_H}{\delta g^{\mu\nu}}
&=&\frac{1}{12}\Big\{
-  \theta^2\Big[ R_{\mu\nu}- \frac12\,g_{\mu\nu} \,R\Big]
- \Big[ g_{\mu\nu}\Box-\nabla_\mu \nabla_\nu\Big] \theta^2
\nonumber\\
&+ &\frac{3\q^2}{2} K \Big[\w_\mu\,\w_\nu- \frac12 \, g_{\mu\nu}\,\w^\rho\,\w_\rho\Big]
-\frac{3\,\q}{2} \Big[ \w_\nu\nabla_\mu+\w_\mu\nabla_\nu - g_{\mu\nu}\,\w^\rho\nabla_\rho\Big]\,K \Big\}
\nonumber\\[4pt]
&+&
\frac12 \, \partial_\mu h \partial_\nu h -\frac14\,g_{\mu\nu} (\partial_\alpha h)^2
+\frac12 \,h^2\,\Big[ \cE_{\mu\nu}-\frac12 \, g_{\mu\nu}\, \cE_{\alpha\beta}\, g^{\alpha\beta}\Big]
+\frac12\, g_{\mu\nu}\, V
\nonumber\\
&-&\frac12\, \Big\{
\Big[
g^{\alpha\beta}\,F_{\mu\alpha} F_{\nu\beta} - \frac14 \, g_{\mu\nu}\,F_{\alpha\beta}\,F^{\alpha\beta}\Big]
+(F\leftrightarrow F^Y)
\nonumber\\
&+&\sin\chi \Big[ g^{\rho\sigma} \big( F_{\mu\rho} F_{\nu\sigma}^Y + F_{\nu\sigma} F^Y_{\mu\rho}\big)
-\frac12\,g_{\mu\nu} F^{\alpha\beta} F_{\alpha\beta}^Y\Big]\Big\}
=0.
\eea

\medskip\noindent
The last two lines are $1/2$ times the stress energy tensor of the
Weyl and hypercharge gauge fields, including their kinetic mixing.
Note that for the full SM action the rhs of this equation should contain
a similar contribution for SU(2) and SU(3) gauge bosons and
a term  $(1/2) T_{\mu\nu}^\psi$  which accounts for the 
stress-energy tensor of SM fermions, which are neglected here.

Taking the trace of (\ref{eqg}):
\bea\label{trace1}
\frac{1}{12}\,\theta^2 R-\frac14 \Box\theta^2 +\frac{\alpha}{4} \omega^\alpha\nabla_\alpha K
-\frac{\alpha^2}{8} \, K \omega_\alpha\omega^\alpha -\frac12\,(\partial_\alpha h)^2
-\frac12 \,h^2 \cE_{\alpha\beta} g^{\alpha\beta}+2 V=0.
\eea
Another form of this equation making obvious its invariance under Weyl gauge symmetry is:
\bea\label{trace2}
\frac{1}{12} \theta^2\tilde R-\frac14\nabla^\mu \Big[\big(\nabla_\mu-\alpha\omega_\mu\big)\theta^2\Big]
-\frac12 \Big[\Big(\nabla_\mu -\frac{\q}{2}\omega_\mu\Big) h \Big]^2
-\frac12 \,h^2\cE_{\mu\nu} \,g^{\mu\nu} + 2 V=0.
\eea
with $\tilde R$  as in (\ref{tildeR}).
After multiplying this equation by $\sqrt{g}$, each term in this equation
is Weyl gauge invariant, hence the whole equation is invariant, as it should.

Further, the equation of motion of $\phi$ that is obtained from the above Lagrangian is
\bea\label{phieq}
\phi^2 \tilde R +\phi^4=0
\eea
This is actually known  ($\phi^2=-\tilde R$), since it was used to linearise the $\tilde R^2$ term.

The equation of motion of $\omega_\mu$ can be written as 
\bea\label{Jmu}
J^\mu+\nabla_\rho \big(F^{\rho\mu}+\sin\chi\,F_Y^{\rho\mu}\big)=0,\qquad\text{where}\qquad
J^\mu\equiv -\frac{\q}{4} g^{\mu\nu}\big(\nabla_\nu - \alpha\omega_\nu)\, K
\eea

\medskip\noindent
Using the anti-symmetry in indices of the Weyl and hypercharge field strengths $F$ and $F_Y$,
by applying the operator $\nabla_\mu$ on the left equation
we find  that the current $J^\mu$ is conserved
\bea\label{ccc}
\nabla_\mu J^\mu=0.
\eea

\medskip\noindent
This result was used in the text, eq.(\ref{J}). For a further  study of  $J_\mu$ see 
Appendix~A in \cite{non-metricity}.

There is also an equation of motion of $h$, but that
brings no new information beyond that of ``trace'' eq.(\ref{trace2}), eq.(\ref{phieq}) and
eq.(\ref{ccc}) of current conservation, because  it is actually a
linear combination of these; hence we replace the equation of $h$  by the much simpler (\ref{ccc}).
Finally, on the ground state, eqs.(\ref{trace2}) and (\ref{phieq}) give
(with $\langle\phi\rangle^2\not=0$) that
\bea\label{relEW}
\langle h\rangle^2=\xi_h/(6\lambda)\,\langle\phi\rangle^2
\eea
 discussed in \cite{SMW} (eq.46). Using (\ref{relEW}), (\ref{b20}) and that
from $\cL_H$ we have $M_p^2=\langle\theta^2\rangle/6$, then
\bea\label{th2}
M_p^2=\frac16\langle\theta^2\rangle=\frac16\,\frac{\langle\phi^2\rangle}{\xi^2} \Big(1+ \frac{\xi^2\,\xi_h^2}{6\lambda}\Big)
  \eea
 which is similar to eq.(\ref{gaugefixing}) in the text.
Using (\ref{b20}), (\ref{B21}), (\ref{relEW}) we also find that
$\Lambda=\langle\phi^2\rangle/4$.
From eq.(\ref{th2}) for $\xi_h/\lambda\ra 0$ which decouples the Higgs,  then
$\Lambda= (3/2) \,\xi^2 \,M_p^2$  as in eq.(\ref{E}). Eqs.(\ref{relEW}), (\ref{th2}) also
apply for the case discussed in the text, see eq.(\ref{Lambda2}) and Section~\ref{b3}.
Finally, from $\phi^2=-\tilde R$ then on the ground
state $R=-4\Lambda$ (assuming $\langle\omega_\mu \omega^\mu\rangle=0$).

\subsubsection{
  SM in integrable geometry (local Weyl symmetry case)}\label{b3}

\medskip\noindent
This case discussed in the text in Section~\ref{s31}
can be  recovered directly from the SMW  (Section~\ref{BBB}):
the action and the equations of motion can be found from
those for the  SMW by setting  $F_{\mu\nu}=0$. For convenience
we present these equations below. First, the quadratic term in $\cL_H$ (\ref{higgsRp})
is linearised with the aid of $\phi$, leading to  eq.(\ref{L2}) with $F_{\mu\nu}=0$.
We do not include in the discussion here the $C_{\mu\nu\rho\sigma}^2$ term.
From eq.(\ref{L2}) $\cL_H$  becomes 
\be
\cL_H= \sqrt{g}
\Big\{ -\frac{1}{12} \theta^2\,R -\frac{\alpha}{4} \omega^\mu \nabla_\mu K
+ \frac{\alpha^2}{8} K\,\omega_\mu \omega^\mu
+ \vert D_\mu H\vert^2- \frac{1}{4} F_{Y\,\mu\nu}^{2} - V\Big\}
\ee

\medskip\noindent
with notation as in (\ref{L3}) to (\ref{hhh}). This gives that
\be\label{omegamu}
\omega_\mu=\frac{1}{\q}\nabla_\mu \ln K,\qquad K=\frac{\phi^2}{\xi^2}+2\,(\xi_h +1)\,H^\dagger H. 
\ee

\medskip\noindent
All previous equations of SMW remain valid if one uses  this value of $\omega_\mu$.
The action becomes
\be
\cL_H= \sqrt{g}
\Big\{ -\frac{1}{12} \theta^2\,R 
- \frac{g^{\mu\nu}}{8} \,\frac{1}{K} \nabla_\mu K\,\nabla_\nu K
+ \vert D_\mu H\vert^2- \frac{1}{4} F_{Y\,\mu\nu}^{2} - V\Big\}
\ee
In the unitary gauge
\be\label{lh}
\cL_H= \sqrt{g}\,
  \Big\{ -\frac{1}{12} \theta^2\,R 
- \frac{g^{\mu\nu}}{8} \,\frac{1}{K} \nabla_\mu K\,\nabla_\nu K
+ \frac12 g^{\mu\nu} \partial_\mu h \,\partial_\nu h
+ \frac12 \,h^2 g^{\mu\nu} \cE_{\mu\nu}- \frac{1}{4} F_{Y\,\mu\nu}^{2}- V\Big\}
\ee

\medskip\noindent
with $K=\phi^2/\xi^2+ (\xi_h+1)\,h^2$. This can also be expressed in terms of final Higgs ($\sigma$)
\be
\cL_H= \sqrt{g}
\Big\{ -\frac{1}{12} \theta^2\,R -\frac12 \,(\partial_\mu\theta)^2
+\frac12\,\theta^2\,\Big(\partial_\mu \frac{\sigma}{\theta}\Big)^2
+ \frac12 \,h^2
g^{\mu\nu} \cE_{\mu\nu}- \frac{1}{4} F_{Y\,\mu\nu}^{2}- V\Big\}
\ee

\medskip\noindent
which recovers eq.(\ref{action3}) in the text. Above  we used
$h=\theta\sinh (\sigma/\theta)$, so $K=\theta^2 \cosh^2 \sigma/\theta$.

The equation of the metric from (\ref{lh}) is a particular case of  (\ref{eqg}):
\medskip
\bea\label{tq1}
\!\!\frac{1}{\sqrt{g}}
\frac{\delta \cL_H}{\delta g^{\mu\nu}}
\!\!\!\!&=&\!\!\!\frac{1}{12}\Big\{
-  \theta^2\Big[ R_{\mu\nu}- \frac12\,g_{\mu\nu} \,R\Big]
- \Big[ g_{\mu\nu}\Box-\nabla_\mu \nabla_\nu\Big] \theta^2\Big\}
\nonumber\\
&- &\!\!\!
\frac18 \frac{1}{K} \,\Big[\nabla_\mu K \nabla_\nu K -\frac12\,g_{\mu\nu} (\nabla_\alpha K)^2\Big] 
+
\frac12 \, \partial_\mu h \partial_\nu h -\frac14\,g_{\mu\nu} (\partial_\alpha h)^2
\nonumber\\[5pt]
&+&\!\!\!\!\frac{h^2}{2}\,\Big[ \cE_{\mu\nu}-\frac12 \, g_{\mu\nu}\, \cE_{\alpha\beta}\,
g^{\alpha\beta}\Big]\!
+\frac{g_{\mu\nu}}{2}\, V
-\frac12\, 
\Big[
g^{\alpha\beta}\,F^Y_{\mu\alpha} F^Y_{\nu\beta}
- \frac{g_{\mu\nu}}{4}\,F^Y_{\alpha\beta}\,F^{Y\,\alpha\beta}\Big]
\!=0.\qquad
\eea

\medskip\noindent
The last bracket is the stress energy tensor of the hypercharge gauge
field. For a full SM action  the above equation should also contain
a similar contribution for the SU(2) and SU(3) gauge bosons and
a contribution $(1/2) T_{\mu\nu}^\psi$  which accounts for the 
stress energy tensor of SM fermions (not included here).
Taking the trace
\bea\label{tq2}
\frac{1}{12}\,\theta^2 R-\frac14 \Box\theta^2
+ \frac{1}{8\,K}\, (\nabla_\mu K)^2 -\frac12\,(\partial_\alpha h)^2
-\frac12 \,h^2 \cE_{\alpha\beta} g^{\alpha\beta}+2 V=0.
\eea

\medskip\noindent
while the equation of motion of $\phi$ gives
\medskip
\be
R-\frac32 \,(\nabla_\mu \ln K)^2 -3 \nabla_\mu \nabla^\mu\ln K=-\phi^2,
\ee

\medskip\noindent
In the lhs of this equation we recognise the expression of
$\tilde R$ for Weyl integrable geometry, see eq.(\ref{tildeR}) in the text with $\w_\mu$ of
(\ref{omegamu}); hence $\tilde R=-\phi^2$,
which we already know from linearising the quadratic term with the aid of $\phi$.

The equation of motion for the Higgs $h$ is simply a linear combination of the last
two equations.  The current $J^\mu$ is in this case trivial, as seen from
using $\w_\mu$ of (\ref{omegamu}) in $J_\mu$ of  (\ref{Jmu}), as also discussed in the text, after eq.(\ref{J}).
Regarding the values of $M_P$ and $\Lambda$, the same
discussion as in (\ref{relEW}) and (\ref{th2}) applies, and again $R=-4\Lambda$ on the ground state.

\vspace{0.4cm}
\noindent
{\bf Acknowledgement:   }
The  work of D.G.  was supported by a grant of the Romanian Ministry of Education and Research,
CNCS-UEFISCDI, project number PN-III-P4-ID-PCE-2020-2255  and partially
by PN 23 21 01 01/2023.

\end{document}